\RequirePackage{fix-cm}
\documentclass{svjour3}                     
\smartqed  
\usepackage{graphicx}
\usepackage{mathptmx}      
\journalname{Applied Physics A}
\begin{document}

\title{Dielectric sensing by charging energy modulation in a nano-granular metal}


\author{Michael Huth \and Florian Kolb \and Harald Plank}


\institute{M. Huth \at Physikalisches Institut\\
Goethe University\\
Max-von-Laue-Str. 1\\
60438 Frankfurt am Main, Germany\\
\email{michael.huth@physik.uni-frankfurt.de}
\and F. Kolb, H. Plank \at Institute for Electron Microscopy and Nanoanalysis\\
Graz University of Technology\\
Steyrergasse 17\\
A-8010 Graz, Austria
}

\date{Received: date / Accepted: date}
\maketitle
\begin{abstract}
Several sensing concepts using nanostructures prepared by focused electron beam induced deposition have been developed over the last years. Following work on highly miniaturized Hall sensors for magnetic sensing with soft magnetic Co, strain and force sensing based on nano-granular platinum-carbon structures (Pt(C)) was introduced. Very recently the capability of nano-granular Pt(C) structures to detect the presence of adsorbate water layers by conductance modulations was demonstrated. For magnetic and strain sensing the underlying physical mechanisms of the sensing effect have been analyzed in detail and are now quite well understood. This is not the case for the adsorbate layer induced conductance modulation effect. Here we provide a theoretical framework that allows
for a semi-quantitative understanding of the observed water-sensing effect. We show how the near-interface renormalization of the Coulomb charging energy in the nano-granular metal caused by the dielectric screening of the polarizable adsorbate layer leads to a conductance modulation. The model can account for the conductance modulation observed in the water adsorbate experiments and can also be applied to understand similar effects caused by near-interface dielectric anomalies of ferroelectric thin films on top of nano-granular Pt(C). Our findings provide a pathway towards optimized nano-granular layer structures suitable for a wide range of dielectric or local potential sensing applications.
%
\end{abstract}
\section{Introduction}
\label{intro}
The potential for the development of highly miniaturized sensor elements based on nanostructures fabricated by focused electron beam induced deposition (FEBID) has been recognized rather recently. Sensitive Hall sensor devices have been developed employing Co-based structures obtained with the precursor CO$_2$(CO)$_8$ with excellent sensitivity for local magnetic stray field detection \cite{Boero2005a,Gabureac2010a}. The development of miniaturized strain and force sensors followed next which took advantage of the nano-granularity of Pt(C)-based FEBID structures obtained from the precursor CH$_3$CpCH$_3$Pt (Cp: cyclopentadienyl) \cite{Schwalb2010a} and proved to be highly interesting for atomic force microscopy sensors \cite{Huth2012a}. The working principle for the exploited strain-resistance effect was found to be rooted in the sensitive dependence of the hopping transport between the metallic Pt nano-grains of about 3\,nm diameter through a carbonaceous matrix \cite{Huth2010a}. Based on the same nano-granular FEBID material the most recent sensor development focused on selective gas sensing using adsorbed water as a model system \cite{Kolb2013a}. In this work a custom-made gas measurement chamber was used to detect the conductance change of Pt(C) FEBID structures in the 5 to 60\,nm thickness range under exposure to different relative humidity levels under well-controlled conditions. For 5\,nm devices conductance modulations above 10\,$\%$ were observed at humidity levels above 40\,$\%$. Only a preliminary analysis of the physical mechanism causing this conductance modulation, namely the influence of the polarizable water adsorbate layer on the inter-granular charge transport, was given in \cite{Kolb2013a}. Here we present a theoretical framework which is able to fully account for the experimental findings. The theoretical treatment is formulated on the mean-field level and can be applied to all transport regimes of nano-granular metals from weak to strong inter-grain tunnel coupling. The formalism is sufficiently general to be applicable to other dielectric sensing scenarios based on three-layer heterostructures with one nano-granular layer. It has already been successfully applied to analyze recent experimental findings on the strong conductance modulations observed in nano-granular Pt(C) structures in close proximity to thin layers of an organic ferroelectricum \cite{Huth2014a}. We can anticipate that our work will be of interest for the area of dielectric or local potential sensing in general and may inspire future work on the basis of highly miniaturized nano-granular sensing layers prepared by FEBID or other thin film techniques.
\section{Modeling approach}
\label{model}
FEBID structures prepared with organometallic precursors often show a nano-granular microstructure, i.~e.\ metallic grains of a few nm of diameter are embedded in a carbonaceous matrix. As long as the metallic grains do not directly touch to form a percolating path through the sample, charge transport is governed by thermally assisted tunneling of the electrons. Recent theoretical work has shown that different transport regimes have to be discriminated depending on the strength of the inter-grain tunnel coupling $g$ and temperature $T$ \cite{Beloborodov2007a}. Here the coupling constant $g$ is a dimensionless quantity measured in units of the spin-degenerate conductance quantum $2e^2/h$. For a material to qualify as a granular metal, the condition $g\ll g_0$ has to be fulfilled. $g_0$ denotes the dimensionless conductance within a grain which is assumed to be in the diffusive limit \cite{Beloborodov2007a}. Due to the small size of the metallic grains, their charge capacity $C$ is very small. Assuming a spherical grain of diameter $2R$ ($R$: radius) the capacitance amounts to $4\pi\epsilon_0\epsilon_r R$, with $\epsilon_0$ and $\epsilon_r$ the dielectric constants of vacuum and the surrounding material, respectively. In the tunneling process each tunnel event is associated with a charging energy which, for one grain in a matrix, is given by $E_C=e^2/2C$. It is the interplay of this charging energy, possible finite size effects for very small grains (smaller than about 1\,nm) and the microstructural disorder in the granular metals that leads to the different transport regimes \cite{Beloborodov2007a}.

In order to develop a model that can account for the change of the conductivity of a thin nano-granular metal prepared in contact with a polarizable medium, such as an adsorbed water layer or a dielectric thin film, the influence of the polarizable medium on the charging energy has to be considered. In principle, also the attenuation length of the electronic wave function at the surface of the grains could be influenced by the presence of the nearby polarizable medium. In this case, the coupling constant $g$ would also be modified. This effect is not considered in the approach followed here, because it will be rather small for the heterostructures considered here. If the matrix were to be replaced fully by a highly polarizable medium, or even a ferroelectric material, polarization-induced changes of $g$ can become relevant, as was recently shown in a theoretical study \cite{Udalov2014a}.

For as-grown Pt(C) nano-granular FEBID structures the temperature-dependent conductivity follows a variant of variable range hopping (correlated VRH) that takes disorder-induced, local background charges and the charging energy of the grains into account \cite{Beloborodov2007a}
\begin{equation}
\sigma(T) = \sigma_0\exp{\left[-(T_0/T)^{1/2}\right]}
\label{eq:sigma_weak}
\end{equation}
with the activation temperature $T_0$ governed by the inelastic co-tunneling range $\xi$ as follows
\begin{equation}
k_BT_0 = \frac{\beta e^2}{4\pi\epsilon_0\epsilon_r\xi} \quad\mbox{and}\quad \xi(T) = \frac{4R}{\ln{(\bar{E}^2/16\pi c_{in}g(k_BT)^2)}}
\label{eq:T0_xi}
\end{equation}
with $\bar{E}=E_C$ and $c_{in}=1$ for simplicity (see \cite{Huth2010a} for details). $k_B$ denotes the Boltzmann constant.

This has been experimentally verified to be the case up to room temperature . By post-growth electron irradiation the tunnel coupling strength between the Pt grains can be strongly increased \cite{Porrati2011a,Plank2011a} up to the point of an insulator-metal transition at a critical coupling strength $g_c \approx 0.3$ \cite{Sachser2011a}. For couplings above $g_c$ the charge transport resembles that of a dirty metal and the dominating correction to the diffusive conductivity value $\sigma_0$ is logarithmic in temperature, as shown in Eq.~\ref{eq:sigma_strong} (three-dimensional case) \cite{Beloborodov2007a}
\begin{equation}
\sigma(T) = \sigma_0\left( 1 -\frac{1}{6\pi g}\ln{\left(\frac{gE_c}{k_BT}\right)}\right)
\label{eq:sigma_strong}
\end{equation}
Within the critical region $g \approx g_c$ theory is not able to provide a conclusive answer to the temperature-dependence of the conductivity.

\begin{figure}
\includegraphics[width=1.0\linewidth]{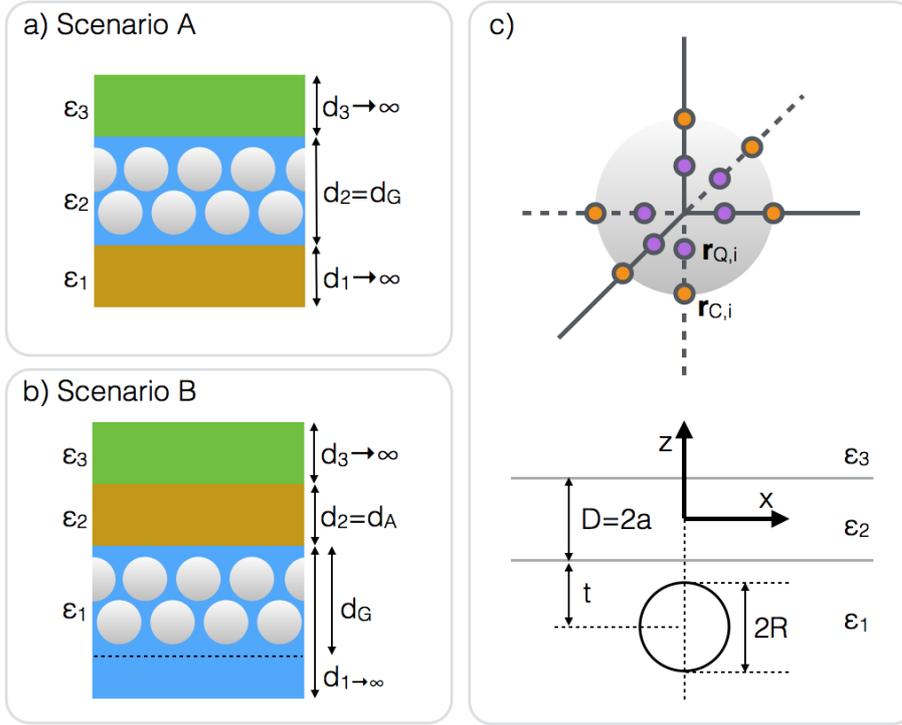}
\caption{Geometric representation of the three-layer heterostructure scenarios (A) and (B). (a) The nano-granular metal layer (2) is sandwiched between the dielectric layers 1 and 3. (b) The nano-granular metal layer forms the bottom layer (1) on top of which a polarizable adsorbate layer (2) and a non or weakly polarizable layer (3) is assumed. (c) One grain of the nano-granular metal together with six point sources at positions $\vec{r}_{Q,i}$ (at half-radius positions) and collocation points $\vec{r}_{C,i}$ (on surface). See text for details. The lower part of (c) represents the geometry for scenario (B) in side view.}
\label{fig:fig1}
\end{figure}

In order to make progress in understanding the observed conductance changes of the nano-granular metal in proximity to an electrically polarizable medium, we introduce the three-layer model structure depicted in Fig.~\ref{fig:fig1}. It consists of a layer with index 2  of thickness $d_2$ and dielectric constant $\epsilon_2$ embedded in two layers of infinite thickness with dielectric constants $\epsilon_1$ and $\epsilon_3$. We wish to calculate the change of the capacitance of a spherical metal grain of diameter $2R$ that represents layer 2 or 1 (see Fig.~\ref{fig:fig1}a or b). The spherical grain is one out of a regular array of grains for which we assume the same diameter. A nano-granular metal layer of given thickness $d_G$ is now formed by stacking several of these arrays in such a way that all neighbored grains are coupled by the same coupling constant $g$. We discriminate two scenarios: (A) The nano-granular metal layer is surrounded by two dielectric half spaces (Fig.~\ref{fig:fig1}a). Half-space 1 represents the substrate and half-space 3 a highly polarizable layer of rather large thickness. In this case any change of the thickness of layer 3 will not lead to changes in the capacitance of a grain in layer 2. (B) The nano-granular metal layer represents layer 1 and is covered by two additional layers. We assume that layer 2 is strongly polarizable, whereas layer 3 could be vacuum or air. In this case we are interested in changes of the grain capacitance, if the thickness of layer 2 changes. This scenario is used here to study the effects of an adsorbed layer, such as water, on the conductance of the nano-granular metal. At this point one may criticize that the second scenario does not take the finite thickness of the nano-granular layer into account and is itself deposited on a different dielectric half-space (substrate). This is a valid point. However, as will be shown next, the three-layer approach followed here allows for a transparent analytic solution of the capacitance problem and is appropriate for describing the adsorbate effects without introducing too much computational complexity. Also, the conceptual steps in this derivation can be extended to more complex layer structures.

We now turn to the capacitance calculation for a spherical grain embedded either in layer 1 or layer 2. Here two steps are needed. In the first step we wish to get a closed expression for the Coulomb potential at any given point $\vec{r}=(x,y,z)$ in layer 1 or 2, if a point charge $Q$ is placed at any other point $\vec{r}'$ in the same layer. Without loss of generality we set $\vec{r}'=(0,0,z')$. In the next step we use this result to calculate the capacitance of a spherical grain employing the source point collocation method \cite{Wasshuber2001a}.
\subsection{Coulomb potential of point charge in three-layer medium}
For the three-layer medium the method of image charges can be used to calculate the Coulomb potential in a straightforward way, if $-a<z'<a$ and $-a<z<a$ (see Fig.~\ref{fig:fig1}a and c) \cite{Anderson1970a}. This is also possible for the case $z'<-a$ or $z'>a$, as was shown in \cite{Barrera1978a}. A substantial reduction of complexity can be reached by Fourier transforming the expressions for the Coulomb potential in $x$ and $y$ ($\vec{R}=(x,y)$). Here we collect the expressions relevant in the present case following \cite{Barrera1978a}:\\
Scenario (A): $-a<z'<a, -a<z<a$
\begin{eqnarray}
V(k;z,z') & = & \frac{Q}{4\pi\epsilon_0\epsilon_2}\frac{2\pi}{k}[ e^{-k|z-z'|} -\Delta\{ L_{12}e^{-k|z+z'+D|}\\ \nonumber
& + & L_{32}e^{-k|z+z'-D|} - 2L_{12}L_{32}\cosh{(k|z-z'|)}\}]
\label{eq:scenA}
\end{eqnarray}
Scenario (B): $z'<-a, z<-a$
\begin{eqnarray}
V(k;z,z') & = & \frac{Q}{4\pi\epsilon_0}\frac{2}{\epsilon_2+\epsilon_3}\frac{2\epsilon_2}{\epsilon_2+\epsilon_1}\frac{2\pi}{k} [e^{-k|z-z'|}\\ \nonumber
& - & e^{-k(|z|+|z'|)} + \Delta e^{-k(|z|+|z'|)}]
\label{eq:scenB}
\end{eqnarray}
with
\begin{equation}
L_{i2} \equiv \frac{\epsilon_i - \epsilon_2}{\epsilon_i + \epsilon_2} \quad\mbox{and}\quad \Delta \equiv \frac{1}{1 - L_{12}L_{32}e^{-2kD}}
\end{equation}
and $D=2a$.

By back transforming into $\vec{R}$-space a closed integral expression for the electrostatic potential is obtained
\begin{equation}
V(\vec{R},z;z') = \frac{1}{2\pi}\int_0^{\infty}k J_0(kR)V(k;z,z') dk
\end{equation}
$J_0(kR)$ denotes the zero order Bessel function of the first kind. This integral has to be evaluated numerically and yields the electrostatic potential for a given point charge distribution which is needed in the capacitance calculation described next. 
\subsection{Capacitance calculation using source point collocation method}
\begin{figure}
\includegraphics[width=0.75\linewidth]{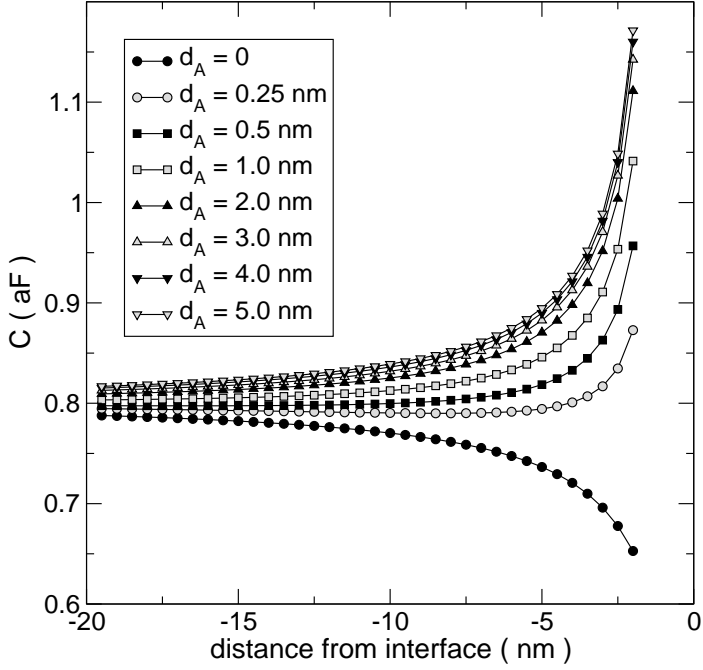}
\caption{Capacitance of spherical grain as a function of the distance from the interface between layer 1 and 2. The nano-granular metal is in layer 1. The different curves refer to different adsorbate layer thicknesses of layer 2, as indicated. See Tab.~\ref{tab:tab1} for the used simulation parameters.}
\label{fig:fig2}
\end{figure}

The source point collocation method for capacitance calculations of arbitrary shaped metallic electrodes in a dielectric medium uses the fact that the surface of a metal electrode is an equipotential surface. In the present case the metallic electrode is a spherical grain and the dielectric medium is formed by the surrounding granular metal which is treated within an effective medium theory according to Maxwell and Garnett \cite{Garnett1904a,Zeng1988a}. This is appropriate within the mean-field treatment of the problem. In order to ensure an equipotential surface on the spherical grain we replace the sphere by $n$ point charges at predefined fixed positions $\{\vec{r}_{Q,i}\}$ within the sphere volume. This is exemplarily shown for $n=6$ in Fig.~\ref{fig:fig1}c. The point charges $Q_i$ have to be determined such that the potential $\phi_0$ on the $n$ collocation points $\{\vec{r}_{C,i}\}$ on the surface of the sphere is constant (see also Fig.~\ref{fig:fig1}c). From this the following system of linear equations results $(j=1,\dots n)$
\begin{equation}
\phi_0 = \sum_{i=1}^n \frac{1}{4\pi\epsilon_0\epsilon_r}\frac{Q_i}{|\vec{r}_{Q,i} - \vec{r}_{C,j}|}
\end{equation}
from which the unknown point charges $Q_i$ are obtained by matrix inversion. The sought for capacitance than follows directly from
\begin{equation}
C = \frac{\sum_{i=1}^nQ_i}{\phi_0}
\end{equation}
In Fig.~\ref{fig:fig2} we show the result of capacitance calculations within scenario (B) performed with $n=6$ and with varying thickness of layer 2, simulating an adsorbate layer. The simulation parameters are listed in Tab.~\ref{tab:tab1}. In all cases we assume a dielectric constant of $\epsilon_2=\epsilon_A=80$ for the adsorbate layer, like for bulk liquid water \cite{Uematsu1980a}. We comment on this simplifying assumption in the discussion section. With regard to the relevant thickness range we use results from scanning probe microscopy (SPM) and dynamic force spetroscopy (DFS) experiments on the coverage of different surfaces with water films under different conditions of relative humidity \cite{Gil2000a,Opitz2007a}. From these experiments we identify the suitable thickness range for $d_A$ to be from 0.25\,nm, corresponding to one monolayer of water, to 5\,nm, as is observed at room temperature for relative humidity levels up to 90\,$\%$. Since the thickness of the water film depends on the substrate material, we took values which were found for graphite \cite{Opitz2007a}. This should be representative for the Pt(C) FEBID structures employed here, if we assume that the surface termination layer of the nano-granular metal is amorphous carbon.
From these calculations we deduce as the most relevant results: (1) With increasing adsorbate layer thickness the capacitance of a spherical grain close to the interface between layer 1 and 2 increases. This effect grows with increasing adsorbate layer thickness and tends to saturate for larger thicknesses. (2) The capacitance increase drops off for larger distances to the interface and tends to saturate for distances above about 20\,nm. From this we can conclude that nano-granular metal layers are most sensitive to this capacitance renormalization effects for thicknesses up to the 20\,nm range.

The accuracy of the capacitance calculation based on the source point collocation method can easily be improved by using $n=8$ or $n=12$. We did comparative capacitance calculations with these larger $n$-values and found a very good correspondence with the $n=6$ calculation within an error margin of less than 4\,$\%$. In the following we use the obtained capacitance values and calculate the corresponding changes in the conductance of a nano-granular metal due to the associated reduction of the charging energy $E_C \propto 1/C$.

\begin{table}
\begin{tabular}{c||c|c|c|c|c|c|c|c} \hline\noalign{\smallskip}
Fig. & $\epsilon_1$ & $\epsilon_2$ & $\epsilon_3$ & $g$ & $T$/K & $R$/nm & $d_A$/nm & $d_G$/nm \\ \hline
2 & 5 & 80 & 1 & --& -- & 1.5 & variable & -- \\
3 & 5 & 80 & 1 & 0.01 & 298 & 1.5 & variable & variable \\
4 & 5 & 80 & 1 & variable & 298 &1.5 & 2 & 22 \\ \noalign{\smallskip}\hline
\end{tabular}
\caption{Table of parameters used for the simulations shown in Figs. \ref{fig:fig2}, \ref{fig:fig3} and \ref{fig:fig4}. For the capacitance calculations $n=6$ was chosen. The fixed source point positions were set to the sphere half-radius value along the three orthogonal axes from the center of the sphere, as is schematically indicated in Fig.~\ref{fig:fig1}c.}
\label{tab:tab1}
\end{table}
In order to calculate the conductance $\sigma$ of a nano-granular metal layer of thickness $d_G$ we employ a simple parallel resistor model. For any given thickness $d_G$ we consider the layer as a parallel circuit of the conductance of $m$ single layers of identical spherical grain arrays. For each grain array, positioned at a certain distance $z_i$ form the layer 1 / layer 2 interface, we use the capacitance of a single grain calculated for this distance. From this we obtain the corresponding single-layer conductance $\sigma(z_i)$ and the overall conductance $\sigma=\sum_i \sigma(z_i)$. In all case we assume that the first layer of grains (next to the interface) is at a distance $z_1 = R+0.5$\,nm. The distance between subsequent single layers is assumed to be $2R$. This corresponds approximately to a close-packed grain arrangement with a 0.5\,nm distance (tunnel barriers) between the surfaces of neighboring grains. The calculated conductance values given below are all stated as relative changes of the conductance for different adsorbate layer thicknesses
\begin{equation}
\Delta\sigma = \frac{\sigma_{d_A} - \sigma_{d_{A,ref}}}{\sigma_{d_{A,ref}}}
\end{equation}
They do not depend sensitively on the chosen geometric model for the microstructural arrangement of the spherical grains.

At room temperature this model is adequate since the average residence time of an excess electron on a single grain is small due to the thermally assisted tunneling. In this case non-homogenous current distributions due to local Coulomb blockade effects can be neglected. Also, in the strong coupling limit $g > g_c$ the parallel resistor model is appropriate. However, additional complexity arises at low temperatures in the weak-coupling limit. In this case a distribution of charges in the system of nano-grains has to be calculated such that the overall electrostatic energy of the system is at its minimum. This has been done as a function of the applied bias voltage in the limiting case of next-neighbor tunneling only \cite{Middleton1993a}. Within the correlated VRH scenario no corresponding theoretical analysis is available.

We now discuss the consequences of the increased capacitance due to the adsorbate layer for the weak- and strong-coupling regime.
\paragraph{Weak-coupling regime}
In the weak-coupling regime, following the correlated VRH transport behavior, the renormalized capacitance of a grain close to the interface leads to two modifications in the respective activation temperature $T_0$. As can be seen from Eq.~\ref{eq:T0_xi}, $T_0$ depends on the inelastic co-tunneling range which, on the other hand depends on the effective dielectric constant of the nano-granular metal $\epsilon_r$ and the charging energy $E_C$. We use the calculated capacitance $C(z_i)$ of a single grain within a grain array at a given distance to the interface to obtain the renormalized dielectric constant $\epsilon_r(z_i)$ and charging energy $E_C(z_i)$ as follows
\begin{equation}
\epsilon_r(z_i) = \epsilon_1\frac{C(z_i)}{C_{G,0}} \quad\mbox{and}\quad E_C(z_i) = \frac{e^2}{2C(z_i)} 
\label{eq:eq_cvrh_EC_epsg}
\end{equation}
with $C_{G,0}$ representing the grain capacitance in vacuum, i.~e.\ $C_{G,0}=4\pi\epsilon_0 R$. From this the corresponding layer-specific activation temperature $T_0(z_i)$ is obtained which yields via Eq.~\ref{eq:sigma_weak} the single-layer conductance.

\begin{figure}
\includegraphics[width=0.75\linewidth]{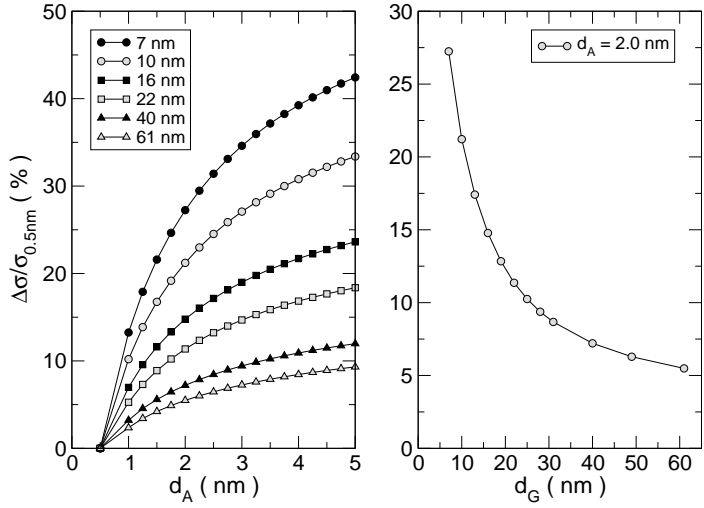}
\caption{(a) Conductance change of nano-granular metal layers of different thickness as function of the water adsorbate layer thickness. As a reference value a two-monolayer coverage, i.~e.\ $d_A=0.5$\,nm, was assumed. (b) Conductance change at fixed adsorbate thickness as function of the nano-granular layer thickness. See Tab.~\ref{tab:tab1} for the used simulation parameters.}
\label{fig:fig3}
\end{figure}

\begin{figure}
\includegraphics[width=0.75\linewidth]{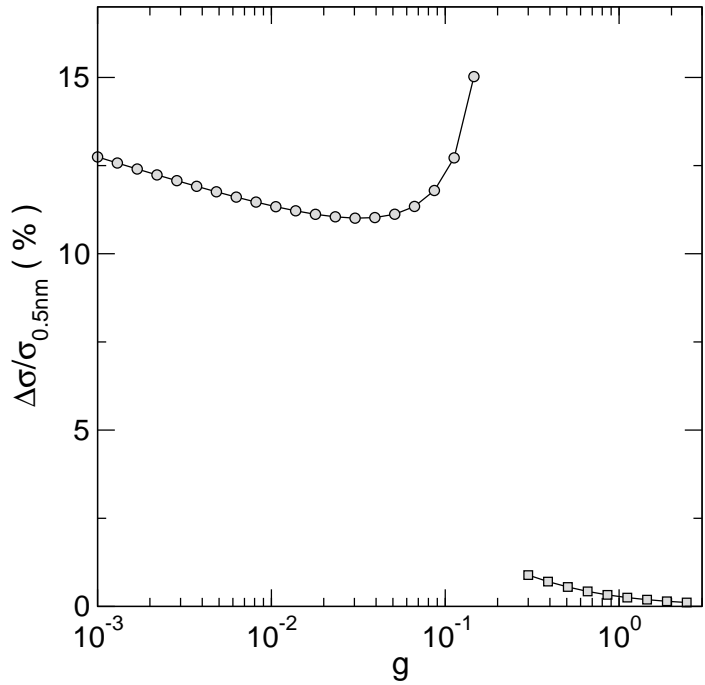}
\caption{Conductance change of a nano-granular metal with fixed values for $d_G$ and $d_A$ as function of the inter-granular coupling strength $g$ within the weak (left) and strong-coupling regime (right). The critical region around $g=g_c$ is shown by the gray area. See Tab.~\ref{tab:tab1} for the used simulation parameters.}
\label{fig:fig4}
\end{figure}

In Fig.~\ref{fig:fig3}a we show the result of a calculation of the relative conductance change of nano-granular layers of different thickness as a function of the adsorbate layer thickness $d_A$. In Fig.~\ref{fig:fig3}b the dependence of the relative conductance change on the layer thickness $d_G$ is depicted for a fixed adsorbate layer thickness of 2\,nm. Apparent qualitative trends are a rather strong conductance modulation of several 10\,$\%$ for a nano-granular metal consisting of only two grain layers ($d_G=7$\,nm) as the adsorbate layer thickness increases to 5\,nm, and a fast drop of the conductance change with increasing thickness of the nano-granular metal. The parameter values for $d_A$ and $d_G$ were selected with a view to previous experimental observations of the water adsorbate effect shown in Fig.~\ref{fig:fig5}. A comparison of our model calculations with these experimental results follows in the discussion section.
\paragraph{Strong-coupling regime}
The model calculations in the strong-coupling regime are governed by the logarithmic correction of the temperature dependence of the conductance (see Eq.~\ref{eq:sigma_strong}). They use the renormalized values for the charging energy $E_C(z_i)$, as already given in Eq.~\ref{eq:eq_cvrh_EC_epsg}. One would expect a rather weak response of the conductance on an adsorbate layer, since in the strong-coupling regime the charging energy is not very relevant anymore. This is confirmed by the results of the model calculations shown in Fig.~\ref{fig:fig4} which presents the dependence of the conductance modulation on the coupling strength in the weak (left) and strong-coupling regime (right). Within the critical region the theory is not applicable. In this respect the apparent increase of the conductance modulation as the critical region is approached from the weak-coupling side should be considered with caution. In the next section we discuss the results of the model calculation with a view to the experimental observations.
\section{Discussion}
\label{discussion}
\begin{figure}
\includegraphics[width=0.75\linewidth]{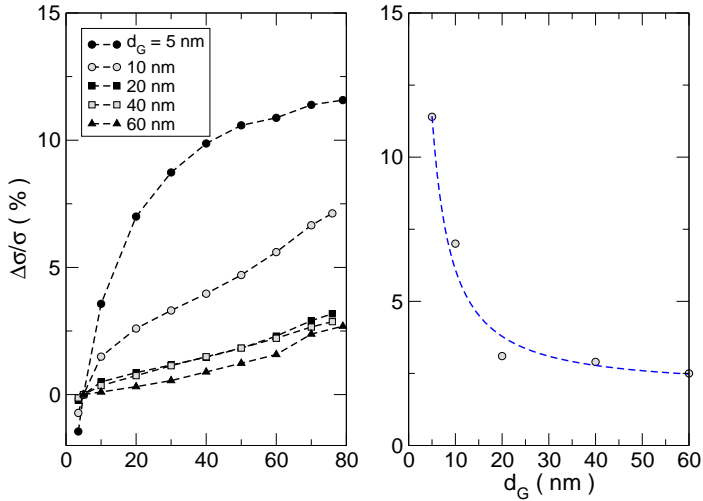}
\caption{(a) Conductance change of nano-granular Pt(C) FEBID layer of different thickness as function of the relative humidity level, i.~e.\ water adsorbate layer thickness. As a reference value the conductance in the presence of a water adsorbate layer at 5\,$\%$ humidity level was used. The temperature was stabilized at 298\,K. (b) Conductance change at fixed relative humidity level of 75\,$\%$ as function of the nano-granular layer thickness. The dashed line represents a guide to the eye.}
\label{fig:fig5}
\end{figure}

In comparing the results of the model calculations with the experiments presented in \cite{Kolb2013a}, several aspects have to be considered. In Fig.~\ref{fig:fig5}a we show the experimental results for the conductance modulation of nano-granular metal layers of different thickness on the relative humidity level. Fig.~\ref{fig:fig5}b depicts the relative conductance change dependence on the nano-granular layer thickness at a fixed relative humidity level. The representation of the data in this form allows for a direct comparison with the model calculations shown in Fig.~\ref{fig:fig3}. The experiments were performed under well-controlled temperature (298\,K) and relative humidity level (5 to 80\,$\%$) conditions, as described in detail in reference \cite{Kolb2013a}. A quantitative comparison of our calculations with the experimental data is not readily possible since, to the best of our knowledge, the relation between the adsorbate layer thickness and the relative humidity level is not known. What can be stated from independent research using scanning probe \cite{Gil2000a} and dynamic force spectroscopy \cite{Opitz2007a} is the following: (1) The water adsorbate layer thickness changes from one monolayer (0.25\,nm) to 5\,nm as the humidity level grows from about 0 (low vacuum conditions) to more than 90\,$\%$ under ambient temperature conditions. (2) The actual layer thickness for a given humidity level may show transient behavior, in particular, a crossover from droplet formation (about 5\,nm height) to closed layers (about 2\,nm thickness) over several hours. (3) The adsorbate layer thickness depends on the substrate material and can vary by almost an order of magnitude between hydrophilic and hydrophobic surfaces.

Depending on the assumed relation between the final adsorbate layer thickness and the relative humidity level, the model calculations can account on a qualitative or even semi-quantitative level for the experimental observations. We note, in particular, that the experimentally observed drop by about a factor of 2.9 (refer to dashed line in Fig.~\ref{fig:fig5}b) of the conductance going from $d_G=5$\,nm to 20\,nm is very well reproduced by the model calculations which yield a factor of 2.4 (cmp. values for $d_G=7$\,nm and 22\,nm). We are therefore confident that our treatment on the mean-field level covers the essential aspects of the physics involved. Nevertheless, we would like to stress again that the exact relationship between the adsorbate layer thickness and the relative humidity level is not known. Furthermore, we would expect that the dielectric properties of the adsorbate layer will differ from the bulk properties in the few monolayer regime. This has not been accounted for in our model, although this could be remedied by a more sophisticated calculation in which the adsorbate layer itself is modeled with a $z$-dependent dielectric function $\epsilon_2(z)$. In this case the capacitance calculations become much more involved and one may need to resort to a full numerical treatment of the ensuing boundary value problem.
\section{Conclusions}
\label{conclusions}
We have presented a mean-field modeling approach that describes the influence of an electrically polarizable medium on the temperature-dependent conductance of a nano-granular metal in the weak and strong inter-granular coupling regime. Based on a three-layer heterostructure geometry two scenarios have been discriminated. In scenario (A) the nano-granular metal is assumed to be sandwiched between two dielectric half spaces. This scenario proved to be successful in describing the strong conductance modulation observed for a nano-granular metal at the paraelectric-ferroelectric phase transition of an organic ferroelectric thin film on top of a nano-granular Pt(C) FEBID structure \cite{Huth2014a}. Scenario (B) takes the nano-granular metal layer as the bottom layer of the three-layer heterostructure, the second layer is assumed to be a polarizable adsorbate and the third layer represents vacuum or any other non or weakly polarizable medium. This scenario was successfully applied to describe the conductance modulations observed for Pt(C) FEBID layers of 5 to 60\,nm thickness covered with adsorbed water under well-controlled relative humidity levels. We can envision a rather wide range of additional sensor applications that use the influence of the polarization properties of different media in contact with thin nano-granular metal layers prepared by FEBID. Due to the excellent down-scaling capabilities of FEBID, spatially resolved sensor structures can be realized. Our findings suggest that the correlated variable range hopping, i.~e.\ weak-coupling, regime is most sensitive to dielectric modulation effects. Assuming a typical inelastic co-tunneling range of about 20\,nm, the working principle introduced here for nano-granular sensor structures should scale down gracefully to the sub-50\,nm range.
\begin{acknowledgements}
M. H. thanks the Deutsche Forschungsgemeinschaft for financial support through the Collaborative Research Centre SFB/TR\,49.
\end{acknowledgements}

\begin{thebibliography}{}
\bibitem{Boero2005a} G. Boero, I. Utke, T. Bret, N. Quack, M. Todorova, S. Mouaziz, P. Kejik, J. Brugger, R. S. Popovic, P. Hoffmann, Appl. Phys. Lett. {\bf 86}, 042503 (2005)
\bibitem{Gabureac2010a} M. Gabureac, L. Bernau, I. Utke, G. Boero, Nanotechnology {\bf 21}, 115503 (2010)
\bibitem{Schwalb2010a} C. H. Schwalb, C. Grimm, M. Baranowski, R. Sachser, F. Porrati, H. Reith, P. Das, J. Müller, F. Völklein, A. Kaya, M. Huth, Sensors {\bf 10}, 9847 (2010)
\bibitem{Huth2012a} M. Huth, F. Porrati, C. H. Schwalb, M. Winhold, R. Sachser, M. Dukic, J. Adams, G. Fantner, Beilstein J. Nanotechn. {\bf 3}, 597 (2012)
\bibitem{Huth2010a} M. Huth, J. Appl. Phys. {\bf 107}, 113709 (2010)
\bibitem{Kolb2013a} F. Kolb, K. Schmoltner, M. Huth, A. Hohenau, J. Krenn, A. Klug, E.J.W. List, H. Plank, Nanotechnology {\bf 24}, 305501 (2013)
\bibitem{Huth2014a} M. Huth, A. Rippert, R. Sachser, L. Keller, submitted (2014)
\bibitem{Beloborodov2007a} I. S. Beloborodov, A. V. Lopatin, V. M. Vinokur, K. B. Efetov, Rev. Mod. Phys. {\bf 79}, 469 (2007)
\bibitem{Udalov2014a} O. G. Udalov, N. M. Chtchelkatchev, A. Glatz, I. S. Beloborodov, Phys. Rev. B {\bf 89}, 054203 (2014)
\bibitem{Porrati2011a} F. Porrati, R. Sachser, C. H. Schwalb, A. Frangakis, M. Huth, J. Appl. Phys. {\bf 109}, 063715 (2011)
\bibitem{Plank2011a} H. Plank, G. Kothleitner, F. Hofer, S. G. Michelitsch, C. Gspan, A. Hohenau, J. Krenn, J. Vac. Sci. Techn. A {\bf 29}, 051801 (2011)
\bibitem{Sachser2011a} R. Sachser, F. Porrati, Ch. H. Schwalb, M. Huth, Phys. Rev. Lett. {\bf 107}, 206803 (2011)
\bibitem{Wasshuber2001a} C. Wasshuber, Computational Single-Electronics, 139ff. Springer, Wien / New York (2001)
\bibitem{Anderson1970a} see e.~g. N. Anderson, Am. J. Phys. {\bf 38}, 1483 (1970)
\bibitem{Barrera1978a} R. G. Barrera, O. Guzman, B. Balaguer, Am. J. Phys. {\bf 46}, 1172 (1978)
\bibitem{Garnett1904a} J. C. Garnett, Philos. Trans. R. Soc. London {\bf 203}, 385 (1904); {\bf 205}, 237 (1906)
\bibitem{Zeng1988a} X. C. Zeng, D. J. Bergmann, P. M. Hui, and D. Stroud, Phys. Rev. B {\bf 38}, 10970 (1988)
\bibitem{Uematsu1980a} M. Uematsu, E. U. Frank, J. Phys. Chem. Ref. Data {\bf 9}, 1291 (1980)
\bibitem{Gil2000a} A. Gil, J. Colchero, M. Luna, J. Gomez-Herrero, A. M. Baro, Langmuir {\bf 16}, 5086 (2000)
\bibitem{Opitz2007a} A. Opitz, M. Scherge, S. I.-U. Ahmed, J. A. Schaefer, J. Appl. Phys. {\bf 101}, 064310 (2007)
\bibitem{Middleton1993a} A. A. Middleton, N. S. Wingreen, Phys. Rev. Lett. {\bf 71}, 3198 (1993)
\end{thebibliography}
\end{document}